\documentclass[12pt]{article}
\usepackage{color}
\usepackage{geometry}
\usepackage{authblk}
\usepackage{amsmath}
\usepackage{graphicx}
\usepackage{setspace} 
\usepackage{cite}
\usepackage{xfrac}

\providecommand{\keywords}[1]{\textbf{\textit{Keywords---}} #1}

\title{Using a (slightly) more realistic model resolves the perfect mixing paradox}

\author[1,2]{Rodrigo Ramos-Jiliberto\thanks{ramos.jiliberto@gmail.com (corresponding author)}}
\author[3,4]{Pablo Moisset de Espan\'es\thanks{pmoisset@ing.uchile.cl}}
\affil[1]{Centro Nacional del Medio Ambiente. Universidad de Chile. Av. Alcalde Fernando Castillo Velasco 9975, La Reina, Santiago, Chile }
\affil[2]{Programas de Postgrado, Facultad de Ciencias, Pontificia Universidad Cat\'olica de Valpara\'iso. Av. Brasil 2950, Valpara\'iso, Chile}
\affil[3]{Instituto de Din\'amica Celular y Biotecnolog\'ia, Av. Beaucheff 850, Santiago, Chile}
\affil[4]{Centro de Modelamiento Matem\'atico, FCFM, Universidad de Chile, Av. Beaucheff 850, Santiago, Chile}

\begin{document}
\maketitle

\begin{abstract}
In a recent paper published in Ecosphere, their authors suggest that extending the logistic growth model in its usual $r-K$ parameterization to a multi-patch environment results in undesirable properties, that were referred to as the ``perfect mixing paradox''. This led the authors to recommend using the Verhulst $r-\alpha$ parameterization of the logistic model instead and abandoning the term ``carrying capacity'' in the context of population dynamics. In this study we show that the use of the logistic equation in its traditional $r-K$ parameterization is appropriate for representing the dynamics of populations in a simple spatial context with passive migration among patches. Furthermore, we show that the conclusions of the mentioned paper depend on the specific functional form for migration rates used for their analyses. In addition, we suggest that their specific migration model was not the best choice since biologically realistic options exist. Moreover, these alternatives require the same number of parameters. The model we present here is free of the paradoxical behaviors presented previously. This allows us to conclude that the logistic growth in its usual $r-K$ parameterization is useful in a spatial context if extended appropriately to a multi-patch scenario and consequently there are no reasons to abandon the concept of carrying capacity. Ecologists should pay great attention when using models in scenarios that are more complex or just different from the ones for which the models were created.
\end{abstract}



\flushbottom
\maketitle
%
%
\thispagestyle{empty}

\section*{Introduction}

In a recent paper, Arditi et al.~\cite{Arditi2016} stated that a proper patch model of population dynamics must obey a basic logical property: ''If two patches are linked by migration, and if the migration rate becomes infinite, the two patches become perfectly mixed among each other, and the system must behave as a one-patch model for the total population.'' To illustrate the issue, they studied the following model:

\begin{equation}
\label{eq:growth}
\begin{aligned}
\frac{dN_1}{dt} &= r_1N_1\left(1-\frac{N_1}{K_1}\right)
+\beta\left(N_2-N_1\right) \\
\frac{dN_2}{dt} &= r_2N_2\left(1-\frac{N_2}{K_2}\right)+\beta\left(N_1-N_2\right),
\end{aligned}
\end{equation}

\noindent where $N_i$ with $i={1,2}$ being population size in patch $i$, $r_i$ the local intrinsic per capita growth rate in patch $i$, $K_i$ the local carrying capacity in patch $i$ and $\beta$ the migration rate constant from and to any patch in the population. Note that each equation is the classical formula for logistic growth plus a term describing migration between patches.

Arditi et al.~\cite{Arditi2016} found that the asymptotic dynamics of system \eqref{eq:growth} in the case of perfect mixing (i.e. with $\beta\rightarrow \infty$) is different from the asymptotic dynamics of the sum of the two populations in isolation (i.e. with $\beta = 0$). In particular, they showed that the equilibrium population size of the system with perfect mixing is different (either larger or smaller) that the sum of equilibrium sizes of the isolated populations. In the limiting but plausible case that the local populations differed in the value of their carrying capacities $K_i$ but not in the values of $r_i$, merging two patches in a single one showed to be always detrimental for equilibrium population size.

Although the analysis is correct, it is valid to ask whether the particular choice for describing migration in \eqref{eq:growth} was the best one for studying such a general ecological phenomenon. Apparently, the choice for migration model in\cite{Arditi2016} was made because of two main reasons: 1) this system was analyzed previously \cite{Freedman77} \cite{DeAngelis1979} \cite{HOLT1985181} \cite{hanski1999metapopulation} \cite{DeAngelis20143087} \cite{Arditi201545} thus it has some tradition within the ecological literature, and 2) Arditi et al.~\cite{Arditi2016} considered this model as a ``natural way'' to represent a two-patch system with logistic growth.

All other things being equal, a well known and widely used model should be favored over its competitors. However this is only valid until we consider a model presenting some objective advantage (e.g. better match with empirical observations) without compromising any substantial aspects (such as number of parameters, mathematical tractability, etc.). In our opinion, model~\eqref{eq:growth} is neither the most natural nor the best way to extend the logistic growth model to a two-patch scenario. Furthermore, we will show below that the paradoxical results reported by Arditi et al.~\cite{Arditi2016} are only a consequence of using the specific model~\eqref{eq:growth} and should not be considered to be a general fact.

\section*{A  more biologically plausible model}

\begin{figure}
\centering
\includegraphics[width=.6\textwidth]{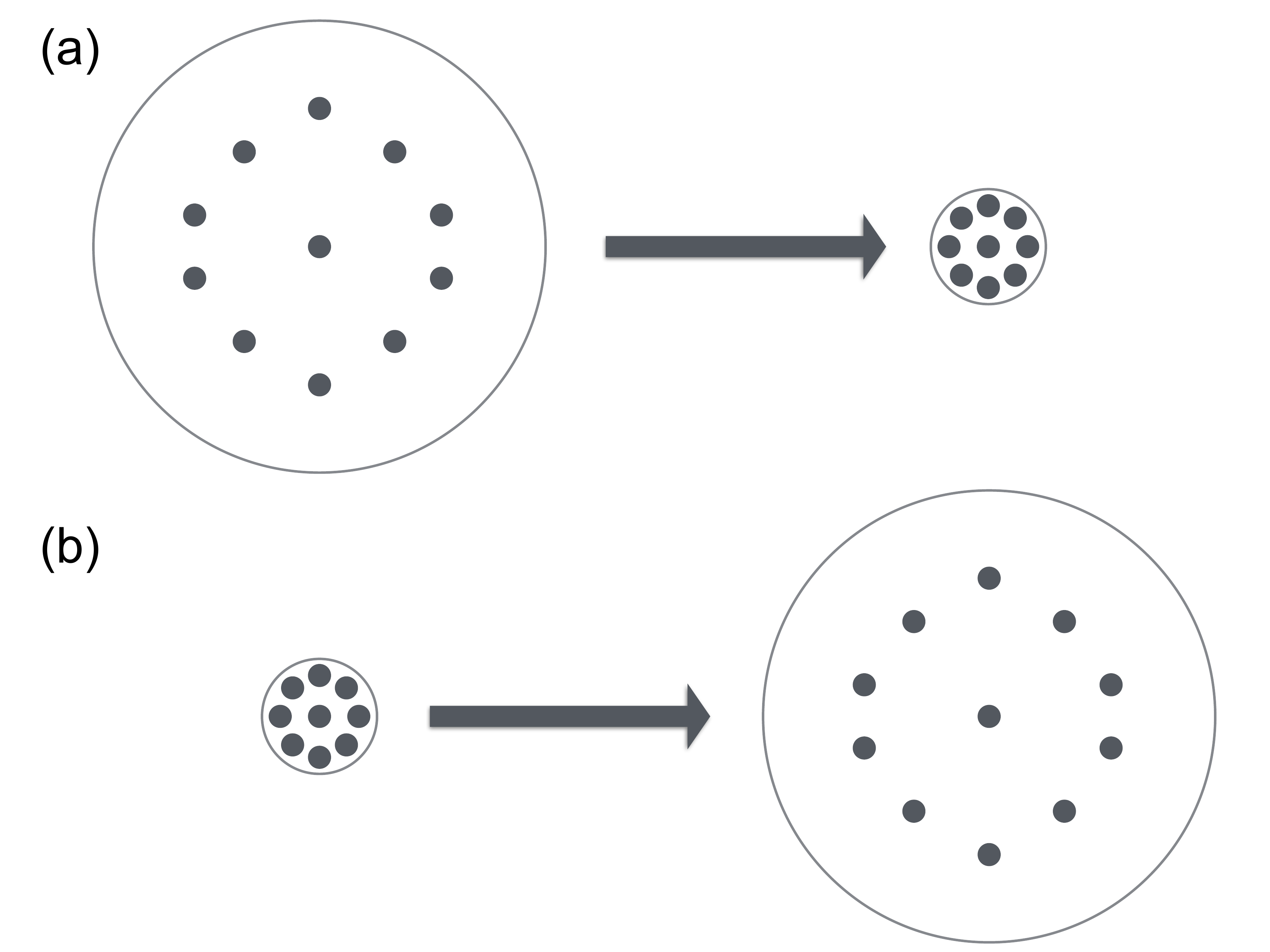}
\caption{\label{fig:fig1}Graphical representation of the flux of migrants in a two-patch population dynamics model. a): biologically unrealistic assumption of model\eqref{eq:growth}, where the net flux of migrants occurs from the less dense (with higher absolute population size but with a much larger patch size) to the denser patch (b): more realistic assumption, with migrant flux from the more dense to the less dense patch}
\end{figure}

Model~\eqref{eq:growth}, used in~\cite{Arditi2016} to present the ``perfect mixing paradox'' contains as a key component a passive migration rate from patch $i$ to patch $j$, $\beta(N_i-N_j)$. This formulation of passive migration rate assumes that there will be a positive flux of migrants from patch $i$ to patch $j$ whenever the absolute population size in patch $i$ is larger than the absolute population size in patch $j$, no matter the differences in patch size or quality. This means that, given equal patch quality, it is possible to have a flux of migrants from a path with greater absolute population size but with lesser population density (with a very large patch size) to a small and more dense patch which possesses a lower absolute population size (Fig.~\ref{fig:fig1}a). This feature of model~\eqref{eq:growth} represents an assumption of limited biological realism. Under the same scenario, a more reasonable assumption is that migrants should pass from the patch with higher population density (absolute population size divided by patch size) to the patch with lower population density (Fig.~\ref{fig:fig1}b).

We propose to re-evaluate the perfect mixing paradox using a slightly different system. This model is both amenable for analysis and contains a more realistic assumption about the direction of the net flux of migrants.

\begin{equation}\label{eq:growth2}
\begin{aligned}
\frac{dN_1}{dt} &=& r_1N_1\left(1-\frac{N_1}{K_1}\right)
+\beta\left(\frac{N_2}{K_2}-\frac{N_1}{K_1}\right) \\
\frac{dN_2}{dt} &=& r_2N_2\left(1-\frac{N_2}{K_2}\right)
+\beta\left(\frac{N_1}{K_1}-\frac{N_2}{K_2}\right)
\end{aligned}
\end{equation}


In this model the flux of migrants is governed by the differences between the ratios $N_i/K_i$. We will refer to the ratio $N_i/K_i$ as the \emph{saturation of patch $i$}, which is a balance between the local population size at time $t$ and the local equilibrium population size $K_i$. The value of $K_i$ depends on the quantity and quality of resources in patch $i$. The direction of the net flux of migrants in this model captures the intuition described in Fig.~\ref{fig:fig1}b. As shown below, this model does not exhibit the paradox presented in~\cite{Arditi2016}.

First, note that in isolation (i.e., with $\beta = 0$), the system converges to ${N_1}^* = K_1,\ \ {N_2}^* = K_2$. This equilibrium is the same as the one of model~\eqref{eq:growth}.
Using the same reasoning used in~\cite{Arditi2016}, if we assume perfect mixing of local populations (i.e. with $\beta\rightarrow\infty$) in
model~\eqref{eq:growth2}, it can be shown that for all $t>0$

\begin{equation}\label{ratios}
\begin{aligned}
\frac{N_1}{K_1}=\frac{N_2}{K_2}
\end{aligned}
\end{equation}

\noindent and therefore, for calculating the saturation of both patches combined:

\begin{equation}\label{ratios2}
\begin{aligned}
\frac{N_1+N_2}{K_1+K_2} = \frac{N_1\frac{K_1}{K_1}+N_1\frac{K_2}{K_1}}{K_1+K_2} = \frac{N_1}{K_1} = \frac{N_2}{K_2}
\end{aligned}
\end{equation}

This shows that total population saturation under perfect mixing is equal to each of the local population saturations.
Now, let us check whether the main paradoxical property presented in~\cite{Arditi2016} holds for our model~\eqref{eq:growth2}. This implies checking whether or not the long term total population size under perfect mixing is equal to total population size in isolation. Adding both equations of system~\eqref{eq:growth2} and using the equalities~\eqref{ratios2} which are valid for the perfect mixing scenario, yields: 

\begin{align}
\frac{dN_T}{dt} = \frac{dN_1}{dt} + \frac{dN_2}{dt} &= (r_1 N_1+r_2 N_2)\left(1-\frac{N_T}{K_T}\right) \nonumber\\
&= \frac{r_1 K_1+r_2 K_2}{K_T}\left(1-\frac{N_T}{K_T}\right)N_T \nonumber \\
&= \bar{r}\left(1-\frac{N_T}{K_T}\right)N_T\label{our N_T}
\end{align}

\noindent where 
$N_T=N_1+N_2$,\ \ $K_T=K_1+K_2$ and $\bar{r}=\dfrac{r_1 K_1+r_2 K_2}{K_T}$.

It is clear that, at equilibrium, the total population size under perfect mixing (i.e. with $\beta\rightarrow\infty$) is $K_T = K_1+K_2$. Thus, using the more realistic model~\eqref{eq:growth2} resolves the main paradoxical behavior presented in~\cite{Arditi2016} for mixed patches. Note also that in the logistic equation for $N_T$ the total intrinsic growth rate $\bar{r}$ is the weighted average of the local intrinsic growth rates, with weights $K_1$ and $K_2$. In the case that the patches differ only in their intrinsic growth rates $r_i$ and do not differ in their carrying capacities (i.e., $K_1 = K_2$), the total intrinsic growth rate reduces to $\bar{r}=(r_1+r_2)/2$. Also, if $r_1=r_2$ then $\bar{r}=r_1=r_2$.

Another issue presented by Arditi et al.~\cite{Arditi2016} is what they call an ``apparent spatial dependency'' of the equation parameters when the dynamics of the total population is represented by the Verhulst equation. The undesirable model property in a multi-patch context is that the value of the self-interference coefficient in the quadratic term decreases with number of patches $S$:

\begin{equation}\label{Verhulst1}
\frac{dN_T}{dt}=\bar{r}N_T-\frac{\bar\alpha}{S} {N_T}^2
\end{equation}

To solve this issue, Arditi et al.\cite{Arditi2016} suggest to treat population size as density, in terms of mean population size per patch $\bar{N}=N_T/S$. When doing so, Eqn.\eqref{Verhulst1} becomes

\begin{equation}\label{Verhulst2}
\frac{d\bar{N}}{dt}=\bar{r}\bar{N}-\bar{\alpha}\bar{N}^2
\end{equation}

\noindent which follows the Verhulst equation. Thus, the form of the equation is invariant in the number of patches in the metapopulation system and their parameters ($\bar{r}$ and $\bar{\alpha}$) are simply the average of the corresponding local patch parameter values.

In our case, and under the same reasoning, considering the average population  in $S$ well-mixed patches, $\bar{N}=N_T/S$, Eqn.~\eqref{eq:growth2} becomes:

\begin{equation}\label{invariant2}
\frac{d\bar{N}}{dt}=\bar{r}\left(1-\frac{\bar{N}}{\bar{K}}\right)\bar{N}
\end{equation}

\noindent with $\bar{K}=K_T/S$. That is, the carrying capacity of the average population is the average of the local carrying capacities. Like Eqn.~\eqref{Verhulst2}, our Eqn.~\eqref{invariant2} is also invariant in the number patches, and their parameters ($\bar{r}$ and $\bar{K}$) are the weighted and aritmetic means, respectively, of the corresponding local patch parameters. Therefore there is no reason to favor the Verhulst's logistic equation over the classical formulation with the familiar $r-K$ parameterization in a multi-patch context, as argued by Arditi et al.~\cite{Arditi2016}.

\section*{Discussion}

The paper by Arditi et al.\cite{Arditi2016} argued that the logistic equation, in its usual $r-K$ parameterization, presents some undesirable properties when used in a multiple patch context. These properties configure what those authors called the ''perfect mixing paradox.'' Arditi et al.~\cite{Arditi2016} also showed that the Verhulst's formulation of the logistic growth model $\sfrac{dN}{dt}=rN-\alpha N^2 $ is less prone to these paradoxical features, as compared with the familiar Lotka formulation $\sfrac{dN}{dt}=rN(1-\sfrac{N}{K})$, when generalized to a multi-patch environment. They conclude, on the basis of the analysis of these models extended to a metapopulation context by including a specific migration function, that the Verhulst formulation should be favored over the Lotka one, and that the term ``carrying capacity'' is misleading and should be abandoned in favor of the more correct ``equilibrium density.''

The supposedly paradoxical behavior of the metapopulation version of the Lotka-Gause model rests, according to Arditi et al.~\cite{Arditi2016}, on two main features that were exemplified considering a two-patch environment as a study case. The first undesirable property is that the total mixed population equilibrium $K_T$ is in general different from the sum of the equilibria in the isolated patches $K_1 + K_2$. This major shortcoming of the analyzed model led Arditi et al.~\cite{Arditi2016} to state that using the term ``carrying capacity'' is incorrect except in specific contexts. The second undesirable feature is the parameter dependence on the number of patches in the system, exhibited by the Verhulst form of the logistic growth model for the total population size. However, when population size is expressed as mean (per patch) abundance the model parameters can be calculated as the average of the local parameters and do not depend on the number of patches. Nevertheless, Arditi et al.~\cite{Arditi2016} claim that this scale-invariance is only exhibited by the Verhulst model and this gives it an advantage over the Lotka-Gause model.

In this paper, we show that the paradoxical behaviors presented by Arditi et al.~\cite{Arditi2016} belong only to the specific variant of the Lotka-Gause model they analyzed. Also, we suggest that the model used by Arditi et al.~\cite{Arditi2016} is not the best choice regarding biological realism. In fact, we present a model as simple as the one they used (two state-variables, five parameters) that is more realistic and is free of the alluded paradoxes exhibited by the Arditi's extensions to both the Lotka-Gause and the Verhulst logistic models. 

The most remarkable advantage of our model \eqref{eq:growth2} is that, unlike both logistic forms used by Arditi et al.~\cite{Arditi2016} in their analysis, total population size at equilibrium of a perfectly mixed metapopulation is equal to the sum of local equilibria. This feature immediately invalidates the criticism posed over the meaning and usefulness of the carrying capacity term. In our model \eqref{our N_T}, global intrinsic growth rate of the metapopulation is not the arithmetic average of  local growth rates but it is equal to the weighted average of the local growth rate parameters. This is very reasonable, since under perfect mixing among patches, the ratios $N_i/K_i$ are equated while their absolute abundances are not. So, it is possible to have patches with contrasting amount of resources (e.g. space or nutrients) and therefore  with unequal population abundances, say $3$ individuals in patch $1$ and $1000$ individuals in patch $2$. Under this scenario, global intrinsic growth rate could not be the arithmetic mean of the local growth rates but it should be closer to the parameter value of the larger population. In the case of the Arditi's model, the absolute population abundances tend to be the same under perfect mixing and so the arithmetic mean and weighted mean are the same. Regarding the second issue stressed by Arditi et al.~\cite{Arditi2016}, we showed that our model \eqref{our N_T} does not suffer from a lack of scale-invariance and that the dynamics of the per patch mean size of the metapopulation is fully consistent with the well known logistic dynamics within a single patch.

In sum, we show here that the criticisms posed by Arditi et al.~\cite{Arditi2016} to the familiar form of the logistic equation attributed to Lotka and Gause are only valid for the particular way in which those authors extended that equation to the multi-patch scenario. We also suggest that their model is not the best choice among other plausible models of the same complexity, and that their criticisms against the usefulness of the carrying capacity as a measure of patch size or richness is not well justified. However, the paper by Arditi et al. has the value of highlighting that modeling population, metapopulation or community dynamics requires more attention than is usually given to and that models should not be applied to any scenario without a rigorous theoretical analysis of their properties.

\section*{Acknowledgments}
This study was supported by FONDECYT grant 1150348.

\bibliography{literature_Arditi}
\bibliographystyle{plain}

\end{document}